\documentclass[aps,prl,showpacs,amssymb,twocolumn,floatfix]{revtex4}
\usepackage{epsfig}

\begin{document}
\title{Hysteresis and competition between disorder and crystallization in sheared and vibrated granular flow}
\author{Karen E. Daniels and Robert P. Behringer}
\affiliation{Department of Physics and Center for Nonlinear and Complex Systems, Duke University, Durham, NC 27708}
\date{\today}

\begin{abstract}
Experiments on spherical particles in a 3D Couette cell vibrated from below and sheared from above show a hysteretic freezing/melting transition. Under sufficient vibration a crystallized state is observed, which can be melted by sufficient shear. The critical line for this transition coincides with equal kinetic energies for vibration and shear. The force distribution is double-peaked in the crystalline state and single-peaked with an approximately exponential tail in the disordered state. A linear relation between pressure and volume ($dP/dV > 0$) exists for a continuum of partially and/or intermittently melted states over a range of parameters. 
\end{abstract}

\pacs{45.70.-n, 
64.70.-p, 
64.60.Cn, 
64.60.My 
}

\maketitle


The relative stability and selection of crystallized and disordered states is fundamental to the study of condensed matter systems. For systems out of equilibrium, the thermodynamic picture under which temperature melts crystalline order is not necessarily valid. Fluctuations provided by external driving, seemingly temperature-like, have been observed to either order or disorder a system. In particular, the mechanisms by which colloidal suspensions shear-melt and shear-order \cite{Ackerson:1981:SIM,Ackerson:1988:SIO}, remain the subject of debate. Granular materials, which are athermal and have strongly dissipative interactions, provide a complementary means to investigate the transition between disorder and crystallization in nonequilibrium systems.

We examine a granular system in which there is a novel phase transition associated with competition between ordering via one type of energy input, and disordering via another. We perform experiments in a variation on the Couette cell, a convenient setup for the study of sheared granular materials \cite{Miller:1996:SFC,Losert:2000:PDS,Mueth:2000:SGM}. An annular region containing monodisperse spheres is vibrated from below and sheared from above, with the mean pressure and volume set from below by a spring. In such a system, the shear and vibration provide competing effects: the system is disordered or crystallized depending on their relative strengths, and the boundary between these two states occurs at equal kinetic energy input from these two driving mechanisms. The freezing transition is hysteretic, similar to a ``freezing by heating'' \cite{Helbing:2000:FHD} transition. In addition, we observe metastable states over a range of packing fractions and forces. These states have the remarkable property that increased internal forces occur for less dense packings, in contrast to conventional mechanical or thermodynamic compressibility. While previous studies have investigated vibrated \cite{Nowak:1997:RIP} or continuously sheared \cite{Tsai:2003:IGD} granular materials separately, the combination of the two leads to novel effects.
                                               

\begin{figure}
\centerline{\epsfig{file=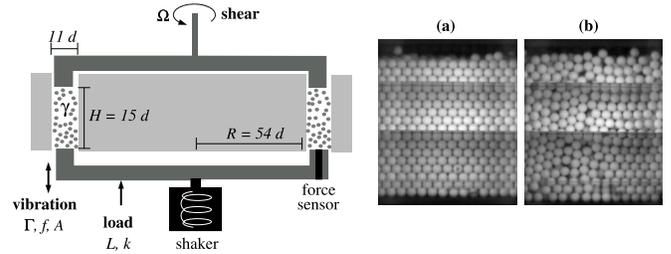, width=\columnwidth}}
\caption{Schematic cross-section of Couette cell experiment (not to scale). Sample images at outer wall, showing crystallized (a) and disordered (b) phases, with glued black particles not visible on upper shearing plate.}
\label{f_exp}
\end{figure}

Fig.~\ref{f_exp} shows a schematic of the apparatus. Monodisperse polypropylene spheres fill the region between two concentric, stationary side walls, with a rotating upper plate attached to a motor and a piston-like bottom plate attached to an electromagnetic shaker with a static internal spring constant $k=341$ N/m. The outer wall is Plexiglas to allow visualization and the bottom plate contains a force sensor flush with its smooth surface. The shear plate surface is a disordered layer of particles. The cell contains 71200 (0.43 kg) polypropylene spheres of diameter $d=2.337$ mm and 5\% polydispersity.

The available parameter space is large: shearing rate $\Omega$ for the upper ring, vibration amplitude $A$ and frequency $f$, peak acceleration $\Gamma \equiv  A (2 \pi f)^2/g$ for sinusoidal vibrations, packing fraction $\gamma$, height $H$, and compressive load $L$. We set $\Omega$ and $\Gamma$ for fixed $f$ (60 Hz) and number of particles; $\gamma$ varies freely. When the system is at its most compact, $L$ is at its lowest; the cell is positioned so that this state has $L_{min} \approx 0$ (zero load) for $\Gamma=0$. We explore shearing rates of $\Omega = 0.01$ to 1.67 Hz (4.2 to 700 $d/s$) and peak accelerations $\Gamma = 0$ to $7$. Even for $\Gamma=7$, the vibration amplitude $A$ is only $0.2d$, small compared to the diameter of the particles. Relative humidity was 31\% to 35\% for all runs; higher humidities shift the transitions observed without affecting the qualitative behavior. We measure forces on a circular plate (diameter $5.4d$) centered on the bottom of the cell. Force measurements are done capacitively, to allow for high speed ($\approx$ kHz) and high sensitivity ($\approx 10^{-5}$ N). The bottom plate moves in response to both the shaker and the particles. A laser position sensor measures the bottom plate position and a piezoelectric accelerometer measures $\Gamma$. Error in the bottom plate position is $\pm 0.03$ mm, and in $H$ is 0.2 mm.


The system crystallizes when sufficiently vibrated, with sustained vibration at $\Gamma \gtrsim 6$ resulting in a hexagonally close packed (HCP) state with a few ($< 10$) grain boundaries and point defects (holes) in the outer visible layer. We refer to this state as {\it crystallized} \cite{bestcrystal}. Examination of the top surface suggests that the order persists throughout the whole cell. Less vigorous vibration results in other HCP orientations and packings in some regions, as well as disordered regions.  As seen in the upper photo in Fig.~\ref{f_exp}, the crystallized state is sufficiently compact that only a few particles are in contact with the shearing surface. In fact, for higher loads, crystallization is suppressed since the $L_{min} \approx 0$ state is no longer accessible.

In contrast, shearing with the upper plate has a disordering effect. The degree of disorder depends on the parameter regime, and the effect is most pronounced near the top of the cell. Running the system at $\Omega=1.61$ Hz and moderate $\Gamma=2.0$ prepares it in a fully {\it disordered} state. At the beginning of each run, we prepare such a state before setting the desired $(\Omega, \Gamma)$. Following such a change, we observe transients of up to five hours. Changes in $H$ are not necessarily monotonic and some states never settle to a characteristic mean $H$, an important effect discussed below.

\begin{figure}
\centerline{\epsfig{file=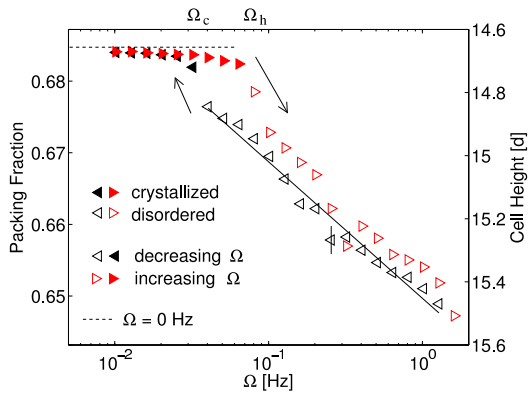, width=0.8\columnwidth}}
\caption{Hysteretic transition from disordered (right) to crystallized (left) state for $\Gamma=2.0$. Dashed line is compaction achieved for $\Omega=0$ Hz, for comparison. The minimum $\gamma$ does not agree with $\gamma_{HCP}=0.74$ due to the curved geometry and error in the measurement of ball and cell dimensions. Vertical bar represents standard deviation of height measurements. }
\label{f_rot_height}
\end{figure}

Fig.~\ref{f_rot_height} shows the hysteretic behavior of transitions between the disordered and crystallized states. As we decrease $\Omega$ in steps from the prepared disordered state, the decrease in $H$ is logarithmic in $\Omega$ \cite{Hartley:2003:LRD}. A first-order phase transition to the crystallized state occurs at $\Omega_c=0.03$ Hz. Further decreases in $\Omega$ result in only slight additional compaction. When we increase $\Omega$ from this crystallized state, the transition to disorder is higher, at $\Omega_h=0.08$ Hz. The shearing causes a transition to less compact state. The total change in $H$ corresponds to $0.8d$, or 5\% of the cell height. While this is a small change in $\gamma$, previous 2D studies \cite{Howell:1999:SF2} indicate that a 4\% decrease in $\gamma$ is sufficient to cross from a state with a strong force chain network to a fluid-like state without force chains.

\begin{figure}
\epsfig{file=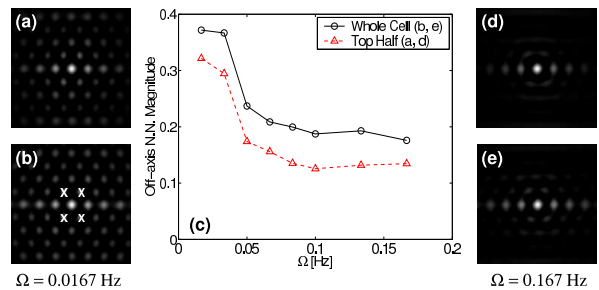, width=0.9\columnwidth}
\caption{(a, b, d, e) Spatial autocorrelation functions (via Wiener-Khinchin method) of outer wall images for $\Gamma=2.0$ and decreasing $\Omega$ starting from disordered state. (c) The average magnitude of the four off-axis nearest neighbor peaks ($\times$) of the autocorrelation functions.}
\label{f_spatcorr}
\end{figure}

For $\Gamma=2.0$, we measured the degree of order as a function of $\Omega$ via 2D spatial autocorrelation functions of images obtained at the outer wall (see Fig.~\ref{f_spatcorr}). The magnitudes of the off-axis peaks, marked $\times$ in (b), change sharply at $\Omega_c$. There is some order present even for $\Omega > \Omega_c$, due to clusters, particularly in the bottom half of cell. Correlations in the direction parallel to the shear remain even as the transverse correlation disappears during the shear melting for $\Omega > \Omega_c$. The remaining longitudinal correlations correspond to linear ``trains'' of particles, and in separate experiments under larger confining pressures have been observed to be the dominant feature \cite{Tsai:2003:IGD}. The importance of clusters in the stability of granular systems has been much investigated \cite{Clusters}, and further work is necessary in this system.

\begin{figure}
\centerline{\epsfig{file=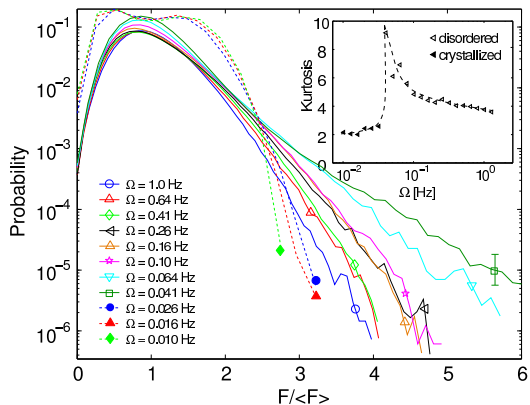, width=0.8\columnwidth}}
\caption{Force PDFs (scaled by mean) for representative decreasing-$\Omega$ ($\lhd$) points in Fig.~\ref{f_rot_height}. Solid lines are disordered and dashed are crystallized. Inset: Kurtosis of the PDFs, showing divergence at $\Omega_c$. Dashed line is a guide to the eye.}
\label{f_forces}
\end{figure}

In addition to the visible changes in $\gamma$ and ordering, there are changes in the interparticle forces as well. Probability distribution functions (PDFs) of the forces measured at the bottom of the cell are markedly different for the two states. As shown in Fig.~\ref{f_forces}a, for $\Omega < \Omega_c$ the PDFs are double-peaked, corresponding to the two peaks in the PDF of a sinusoid. Thus, we infer that the crystallized particles respond to the motion of the bottom plate as a solid body. Time traces of force data with $\Omega>\Omega_c$ show amplitude disorder in the force response, but a well-defined frequency, $f$. The corresponding PDF is single-peaked and the tail of the distribution decays approximately exponentially, a shape characteristic of granular materials \cite{Miller:1996:SFC,Mueth:1998:FDG}. The PDFs for increasing and decreasing $\Omega$ show the same transition, but at $\Omega_h$ and $\Omega_c$, respectively. 

As measured from the kurtosis, the PDF is increasingly wide as $\Omega \rightarrow \Omega_c$ (see Fig.~\ref{f_forces}). This is reminiscent of the large fluctuations near a jamming transition in sheared colloids \cite{Lootens:2003:GSF}, the broadening of the force PDF for simulations of sheared Lennard-Jones particles \cite{OHern:2001:FDN,Snoeijer:2004:FNE}, and the divergence of the viscosity in classical glass transitions. While a crystallized state is fundamentally different from a jammed one due to the presence/absence of order, they share a lack of freedom to rearrange.

As is common with sheared granular materials, the system exhibits an approximately exponential velocity profile localized to the shearing wheel, with a decay length of a few ball diameters. This effect is most extreme in the crystallized state, where the system exhibits stick-slip behavior for the few  balls in contact with the shearing wheel, similar to that observed in granular friction experiments \cite{Nasuno:1998:TRS}. Even in the disordered state, the vibrations do not create a more liquid-like state characterized by an approximately linear velocity profile.

The autocorrelation functions in Fig.~\ref{f_spatcorr} show that disorder occurs throughout the cell for $\Omega > \Omega_c$. This full-depth disordering stands in contrast to the disordering localized to the shear surface in molecular dynamics simulations of a 2D Couette cell with constant-$P$ boundary conditions and gravity, but without vibration \cite{Thompson:1991:GFF}. Since these simulations showed no disordering below the shear band, it appears that vibrations actually augment the disordering process. The effect we observe could be due to either the fluctuations induced by the vibration itself, or the periodically increased contact with the shearing surface. An initially crystallized state sheared without vibrations fails to disorder the lower portion of the cell, indicating that melting is suppressed. The vibrations therefore act subtly: promoting crystallization at low $\Omega$, but also allowing low-shear areas of the cell to disorder at high $\Omega$.

\begin{figure}
\centerline{\epsfig{file=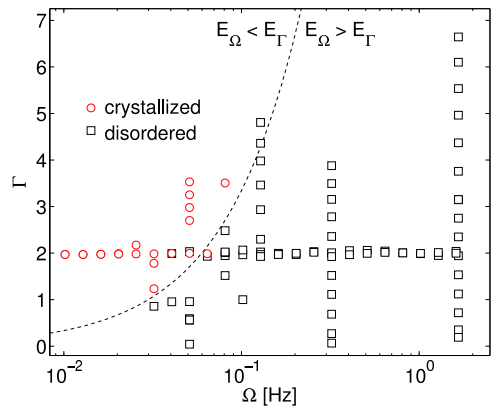, width=0.75\columnwidth}}
\caption{Phase diagram in $\Omega$ and $\Gamma$. Observations of crystallized ($\bigcirc$) and disordered state ($\square$). Overlapping circles and squares are hysteresis from Fig~\ref{f_rot_height}. Dotted line is $E_\Omega=E_\Gamma$.}
\label{f_phase}
\end{figure}

The crystallization/disordering transition can be understood in terms of the relevant energy scales. Ignoring the effects of a compressive pressure ($L_{min}$ is small at $\Omega_c$), shearing and vibration are the only two competing effects. For each, there is an associated kinetic energy input for a particle of mass $m$: $E_{\Omega} = \frac{1}{2} m(R\Omega)^2$ and $E_{\Gamma}= \frac{1}{2} m(2 \pi f A)^2 $. (These two modes are also anisotropic: the shear acts perpendicular to gravity and the vibration parallel.) For $E_\Gamma > E_\Omega$ the system is crystallized and for $E_\Gamma < E_\Omega$ the system is disordered.  The agreement of this line with experimental observations is shown in Fig.~\ref{f_phase}. Hysteresis occurs because the pressure at the line $E_\Gamma = E_\Omega$ depends on the height of the cell.

\begin{figure}
\centerline{\epsfig{file=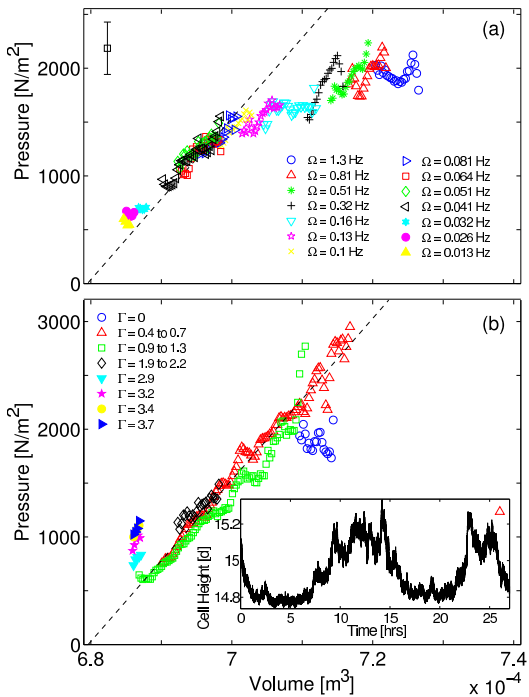, width=0.8\columnwidth}}
\caption{Dependence of ${\bar P}$ on $V$ for (a) $\Gamma=2.0$ and varying $\Omega$ and (b) $\Omega=0.051$ Hz and varying $\Gamma$. Solid symbols are crystallized states. Dashed line is fit to (b), shown also on (a) for comparison. Inset: Intermittency in run marked $\vartriangle$ in (b). }
\label{f_fxcompare}
\end{figure}

The pressure $P(t)$ on the force sensor shows significant fluctuations, but the average $\bar P(V)$, calculated by binning the data by volume $V$, is a smooth function (see Fig.~\ref{f_fxcompare}). Except at the highest $\Omega$ and $\Gamma$, the data for non-crystallized states is consistent with a single value for $dP/dV$, including during transients or in runs with higher $L_{min}$ or higher humidity. At the highest values of $\Omega$, Fig.~\ref{f_fxcompare}a shows a departure from the linear relationship (smaller $dP/dV$), coinciding with shorter tails in the high $\Omega$ force PDFs. Significantly, $dP/dV$ is positive, corresponding to a negative compressibility.

In several experimental runs, the system intermittently exhibited a wide range of partial disorder/crystallization while still remaining along this characteristic line. These runs are the open triangles and squares in Fig.~\ref{f_fxcompare}b and inset. The system can approach arbitrarily close to the fully crystallized state before disordering (dilating), but can also permanently end up in the fully crystallized state. The system displays similar intermittency when started from an over-dilated state and allowed to evolve to a steady state: during net compaction, the system routinely re-dilates slightly before compacting further.

It is remarkable that at lower $\gamma$ the system sustains higher forces, in contrast to the classical mechanical or thermodynamic $\partial P/\partial V < 0$. This effect is a natural consequence of Reynolds dilatancy: shearing causes an increase in $V$. One possibility for the corresponding increase in $P$ is that the larger forces are a sign of well-developed force chains, unmeasured in this experiment. As such, changes in the internal granular pressure would depend not only upon the volume, but also upon an order parameter, $X$, describing the strength of the force chains present. Thus we could write
$$
\frac{dP}{dV} = \frac{\partial P}{\partial V} 
+ \left( \frac{\partial P}{\partial X} \right)_V \frac{d X}{dV}.
$$
Setting a smaller initial $V$ produces a higher $\bar P$, indicating that $\partial P / \partial V$ is negative as for ordinary materials. In 2D sheared systems at constant volume, such as \cite{Howell:1999:SF2}, shearing increased the strength of the force chains and hence the pressure, indicating that the quantity $(\partial P / \partial X)_V$ must also be positive. Therefore, $dX/dV$, while difficult to measure, must also be positive in order to obtain $dP/dV > 0$ as observed in the experiment. This formalism suggests two possible interpretations for the trend in Fig.~\ref{f_fxcompare}a, where for high $\Omega$ $dP/dV$ is positive but decreasing. First, there may be an upper limit to $X$ due to the number of particles available to participate in force chains. Second, if $V$ grows too large, the lateral stability of force chains is compromised. 


In this Letter, we have described competing disordering and crystallizing effects for sheared/vibrated granular materials. While the concept of a granular temperature has provided a thermodynamic analog for granular gases, the findings here point to several caveats regarding that association. External vibration fails to provide ``heating'' in a thermodynamic sense: the resulting fluctuations do not allow the shear profile to be liquid-like, and they produce crystallization rather than melted disorder. However, fluctuations introduced by the shearing of the material have the opposite effect. 
Future work should be undertaken to examine temperature-like variables which might control this phase transition.

Granular materials are often studied with constant-volume constraints, where they exhibit higher forces for higher packing fractions. However, the relationship between $P$ and $V$ appears to involve not only the classical $\partial P/ \partial V < 0$, but a second, opposing, effect dependent upon the internal structure. As a result, lower $\gamma$ states exhibit larger forces. $dP/dV$ is found to have a nearly universal value for a wide range of conditions, suggesting that predictions for its value should be possible. In addition, there is an unexplained logarithmic dependence in $H(\Omega)$ for $\Omega > \Omega_c$. 

The authors would like to thank Lou Kondic, Jerry Gollub, and Matthias Sperl for fruitful conversations. This work has been supported by the NASA microgravity program, grant NNC04GB08G.


\begin{thebibliography}{20}
\bibitem{bestcrystal} Interestingly, the degree of crystallization improves when weak shear ($\Omega \approx 10^{-2}$ Hz) is applied in addition to vibration.
\end{thebibliography}
\end{document}